# Optimization of Electrostatic Quadrupole Profile to Reduce Image Charge Effects

**Selcuk HACIÖMEROĞLU**

Center for Axion and Precision Physics Research, IBS, Daejeon 34051, Republic of Korea

**Abstract:** Vertical electric fields need to remain orders of magnitude smaller than the horizontal electric field in storage ring electric dipole moment experiments. Otherwise, the coupling with the magnetic dipole moment dominates the spin precession, eventually leading to a false signal. This work presents first quantitative studies regarding electrostatic quadrupole profile optimization, which helps suppressing image beam induced vertical electric fields. Besides the demonstration of the profile optimization, its suppression performance is also studied for various beam offsets and beam size configurations. It is found that with the optimum profile, the vertical electric field can be reduced by two orders of magnitude.

## 1. Introduction

Storage ring electric dipole moment (EDM) experiments aim to measure the EDM of proton [1] and deuteron [2] with $d_p \approx 10^{-29}\, e \cdot \mathrm{cm}$ sensitivity. They are designed to measure the vertical spin component of longitudinally polarized beams inside storage rings. Durıng the beam's storage inside the ring with the presence of electric and magnetic fields, its EDM couples with the main field, leading to an out-of-plane spin precession. Ideally, a nonzero measurement indicates a coupling with EDM and the main field.

The major systematic errors in those experiments originate from the magnetic dipole moment (MDM), which is orders of magnitude larger than the EDM. Hence, under certain conditions, even tiny fields (as low as atto-Tesla level magnetic field) can couple with the MDM, and dominate the EDM signal [3]. The systematic error sensitivity of the experiment is mainly determined by how well this effect can be reduced. While a certain reduction can be achieved by means of measurement and compensation, an inhibitory design feature in the experiment is usually more preferable.

According to the T-BMT equation [4, 5], in the presence of only electric field $\vec{E}$, the spin precession is given as

$$\frac{d\vec{s}}{dt} = -\frac{e}{m}\vec{s}\times\left[\left(G+\frac{1}{\gamma+1}\right)\frac{\vec{\beta}\times\vec{E}}{c} - \frac{\eta}{2}\left(\frac{\vec{E}}{c} - \frac{\gamma}{\gamma+1}\frac{\vec{\beta}\cdot\vec{E}}{c}\vec{\beta}\right)\right], \tag{1}$$

where $\vec{s}$ is the spin vector, $c$ is the speed of light, $e$ and $m$ are the electric charge and the mass of the particle, $G$ is the magnetic anomaly (i.e. MDM coupling term), $\eta$ is the EDM coupling term, $\gamma$ and $\vec{\beta}$ are the relativistic Lorentz factor and velocity, respectively. Coupling between $\eta$ and the electric field in the second term determines the EDM signal, while the first term is a source of a false EDM signal.

For particles with a positive $G$, the in-plane precession due to first term can ideally be aligned with momentum by means of the "frozen spin method" [6, 7] with a specific momentum requirement. For proton, which has $G \approx 1.8$, that momentum is 0.7 $\mathrm{GeV}/c$ and it corresponds to $\gamma = 1.248$ and $\beta = 0.59$. Even in the presence of a momentum spread, to first order, every particle can be forced to have that specific momentum by means of RF bunching [8]. However, second order effects like transverse oscillations of the beam around the design orbit still cause a non-negligible spin precession, which can be fixed by means of sextupoles [9, 10]. Therefore, one can safely assume a longitudinal beam polarization during storage time ($\|\vec{s}\| \approx s_l$). The transverse components of the velocity are also expected to be negligible compared to the longitudinal ($\|\vec{\beta}\| \approx \beta_l$). Finally, assuming a mainly radial electric field ($\|\vec{E}\| \approx E_r$), the vertical component of Equation (1) simplifies to

$$\frac{ds_v}{dt} = \frac{es_l}{mc}\left[\left(G+\frac{1}{\gamma+1}\right)\beta_l E_v + \frac{\eta E_r}{2}\right], \tag{2}$$

Here $v$, $l$, and $r$ represent the vertical, longitudinal and radial directions, respectively. As mentioned above, the second term results in the EDM signal, while the first term is a source of systematic error, which must be suppressed.

For the proton EDM experiment parameters with $E_r = 5$ MV/m, in order to suppress the systematic error, average vertical electric field must be $E_v^{avg} < 5$ mV/m. The vertical electric fields can originate from external sources, or misalignments of the ring elements. In many cases, effects from those fields cancel due to symmetries of the ring. In some cases, they need to be measured and cancelled. Another source of the vertical electric field is the image charges at the surrounding material [1], including the vacuum chamber, quadrupole elements, sextupole elements, and so on. For most of the ring elements, including magnetic quadrupoles and sextupoles, vertically oriented parallel plates can shield the electric fields from the vertical image charges[2]. However, limited apertures of electrostatic quadrupoles and sextupoles do not allow such a setup. Therefore, as the quadrupoles occupy 0.1% of the storage ring circumference [1], the vertical electric field at those sections must be kept at $E_v < 5$ V/m.

In this work, we propose an alternative method, namely an optimized quadrupole profile to reduce $E_v$ significantly inside electric quadrupoles, and provide first numerical estimations regarding several experimental conditions, such as beam size, beam offset, and so on. We also demonstrate the optimization of the quadrupole profile.

## 2. Material and Method

Optimization of the electric quadrupole profile and related estimations are conducted numerically by finite difference method, using the 2-dimensional cross section as seen by the beam. Despite being beyond the scope of this works, the method can easily be applied for sextupole profile optimization as well. We basically locate a charge at certain points inside the calculation region, and calculate the image charge induced vertical electric field at those locations.

### 2.1. 2. Estimation of the electric field inside a grounded quadrupole profile

Figure 1 shows the cross section of a vertically off-center, 1 mm radius beam inside a grounded quadrupole-like profile, and the electric potential around it as calculated by solving the Poisson's equation. White and black colors represent the ground and maximum potential, respectively.

The square shaped surface that includes the profile and the inner area (vacuum) is divided into 151 ×151 meshes. Each mesh has 1mm × 1mm dimensions. The shapes of the top/bottom and right/left edges of the quadrupole electrodes are determined by $Y = \pm(a_0 + a_2 x^2)$ and $X = \pm(b_0 + b_2 y^2)$, respectively, where $(x, y)$ is the mesh position with respect to the center of the square, and $a_i$ and $b_i$ are constants that define the curvature and the aperture. In all of the simulations, $b_0 = 4$ cm ($a_0$ and $b_0$ are shown in Figure 2).

The region of interest (gray area) is surrounded by quadrupole plates (shown as white), which are fixed at zero potential. The edges of the square are also fixed at zero potential to avoid floating boundaries. The electrode shapes are determined by $a_i$ and $b_i$ in such a way that the electric potential on them produces quadrupole electric fields at the region of interest. Despite this specific choice, this method works for a wide range of $a_i$ and $b_i$ values as long as the profile does not have too flat or sharp curves. The beam is defined in all simulations by fixed potential meshes (shown in black) with a total of $10^9$ protons. Regardless of the beam size, this region contains as small number of meshes as possible to obtain image charge induced potential distribution at (almost) every point.

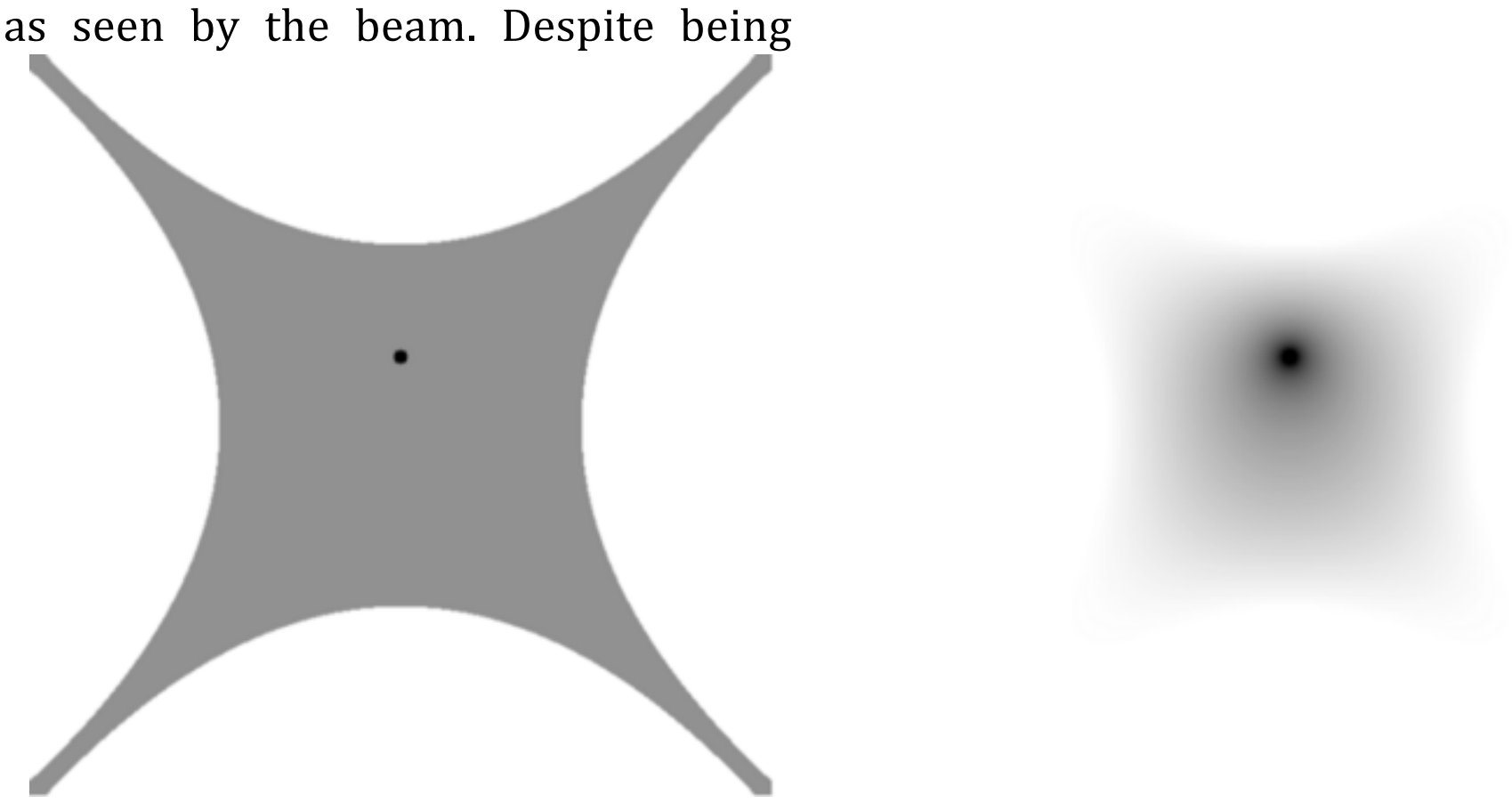

**Figure 1.** *Left:* Cross-sectional drawing of the quadrupole profile (shown in white) and a beam (shown in black) inside. The gray area is vacuum. The electric potential is not shown on this figure. *Right:* Electric potential within the profile is given, as estimated from Poisson's equation. The potential values span the gray tones between white (ground) and black (maximum).

[1] First estimated by William M. Morse.

[2] Proposed by Yannis K. Semertzidis.

The potential distribution around the beam is determined by solving the Poisson's equation numerically, by means of finite difference method. At every point within the boundaries, it is given by the average potential of the surrounding points:

$$V_{i,j} = \frac{1}{4}\left(V_{i-1,j} + V_{i+1,j} + V_{i,j+1} + V_{i,j-1}\right), \qquad (3)$$

where $i$ and $j$ are the mesh indices that are mapped from $x$ and $y$, respectively. It is calculated iteratively on every mesh until the average change between iterations becomes insignificant (<1 ppm). Then, the average vertical electric field on the beam is calculated by integrating the gradient of the electric potential within the beam coverage.

A beam does not experience a vertical electric field from a symmetric quadrupole profile if it is at the geometrical center. On the other hand, the top/bottom electrodes cause vertical, and right/left plates cause both horizontal and vertical forces on the beam, which can be attributed to the asymmetrically distributed image beams. Figure 2 shows the image beams and the corresponding electric forces. It will be shown below that the vertical electric field/force can be cancelled significantly with the correct choice of the aspect ratio $a_0/b_0$.

Finally, let's assume that as in the proton EDM experiment proposal [1], the charged particle beams and the quadrupole elements (with $2a_0 \approx 2b_0 = 8$ cm aperture) have comparable longitudinal length (at the order of 50 cm). In such a configuration, the potential due to the image beams on a plate is approximately 1.5-2 times the 2-dimensional (2D) solution. As the difference is insignificant, the conclusions of these 2D simulations can be generalized to a realistic 3D scenario.

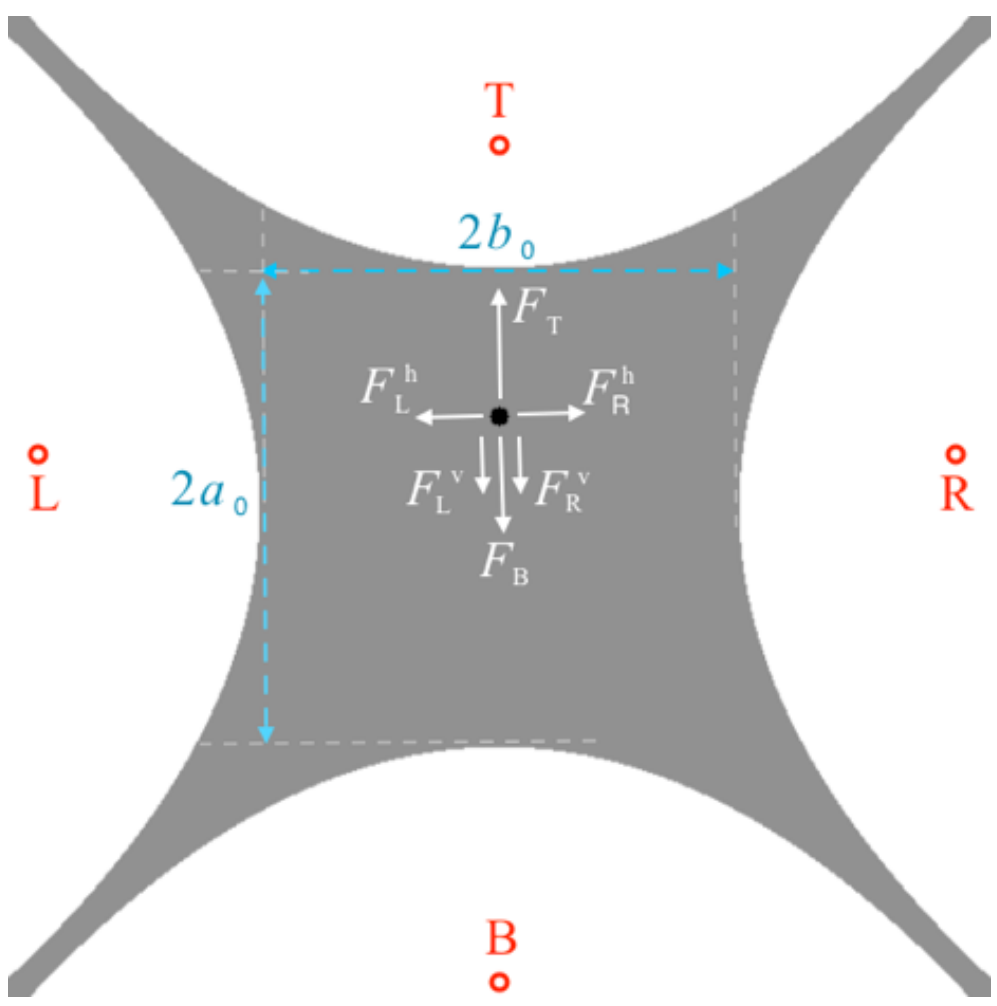

**Figure 2.** Depiction of the simulated setup. The beam is shown by the black circle. The red circles at every electrode, whose locations are not necessarily accurate in this drawing, represent the image beams. They apply a net vertical electric field/force on the beam. As will be shown below, the vertical field can be cancelled by a correct choice of $a_0/b_0$.

## 3. Results

Figure 3 shows the vertical electric field on the beam due to the image beams as a function of its vertical offset. The simulations are made for a beam of 5 mm radius and $10^9$ particles. The aspect ratio of the profile is $a_0/b_0 = 1$. The slope of the line indicates that the image beams apply a 180 V/m net vertical electric field per 1 mm vertical beam offset. Then, following the conclusions of Equation 2, the vertical beam offset must be kept smaller than 25 μm to avoid related systematic errors.

Clearly, the effect becomes more significant for larger beam currents, eventually requiring a beam and quadrupole alignment with a few micrometers of precision. As will be shown below, beam width also has an enhancing effect. It is worth emphasizing that the vertical offset can originate from external fields as well as quadrupole misalignments. Therefore, suppressing the effect rather than monitoring the misalignments is a more robust solution.

As mentioned above, different quadrupole profiles can produce different vertical electric fields. This behavior originates from the balance between the side plates and the top and bottom plates (See Figure 2). Figure 4 shows the net vertical electric field on the beam as a function of the aspect ratio ($a_0/b_0$). The beam is located 1 mm above the geometrical center. As $a_0/b_0$ approaches to 1.7, the vertical electric field components due to the image charges cancel each other out and $E_v$ crosses through zero. Then, it starts growing in the opposite direction.

In the proton EDM experiment proposal, the proton beam oscillates around the design orbit with approximately 1 cm amplitude. Figure 5 shows the vertical electric field on the beam as a function of vertical beam offset $y$ within ±1 cm range. The beam radius is 5 mm, and the aspect ratio is kept at 1.7 in the simulations. As expected, irrespective of the aspect ratio, $E_v \to 0$ as $y \to 0$. For nonzero $y$ values, $E_v$ remains roughly two orders of magnitude smaller than the $a_0/b_0 = 1$ case of Figure 3. Similar to the $a_0/b_0 = 1$ case, $E_v$ has opposite sign at positive and negative $y$ positions. The zero crossings at around ±7 mm could be due to numerical error. Despite being less clear because of the scale, the same wavy behavior is visible in Figure 3, as well.

One concern for the cost-effective production of the quadrupoles is the machining precision. For a quadrupole with $b_0 \approx 4$ cm, $a_0$ differs by roughly 1.2 mm between $a_0/b_0 = 1.67$ and $a_0/b_0 = 1.7$ profiles. That is, machining a profile within $1.67 < a_0/b_0 < 1.7$ requires 0.6 mm accuracy, which is easily achievable with the current technology.

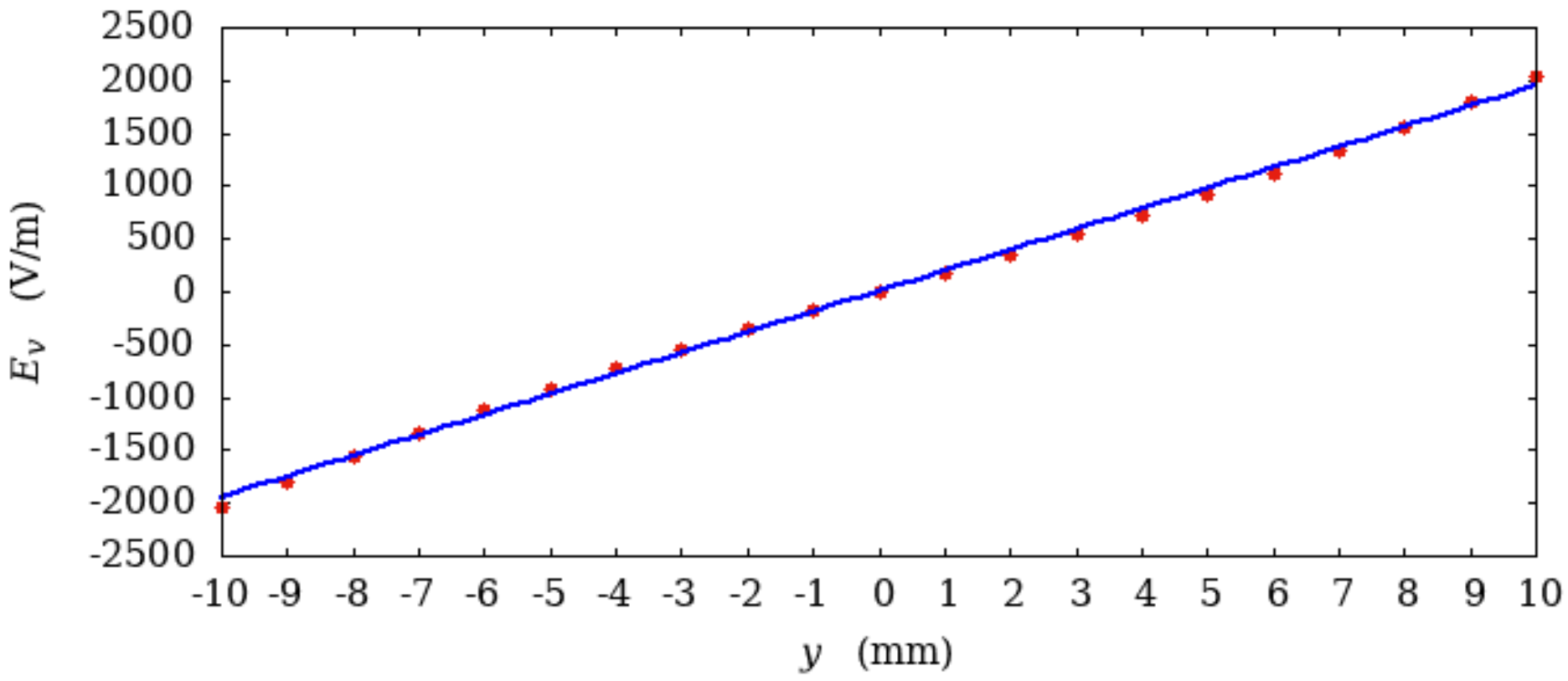

**Figure 3.** Average vertical electric field ($E_v$) on the beam inside a quadrupole profile with $a_0/b_0 = 1$, as a function of vertical offset. The vertical electric field that originates from the image beams on the quadrupole electrodes grows linearly with the vertical position.

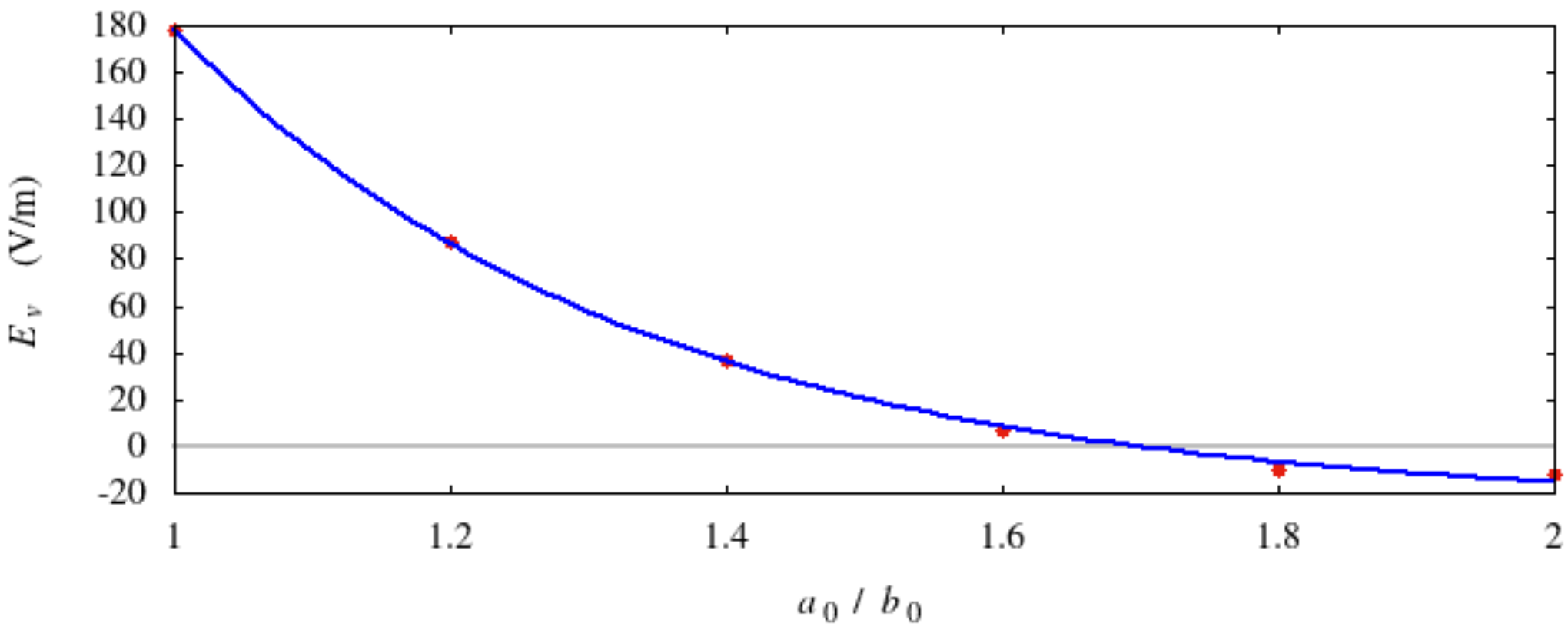

**Figure 4.** Net vertical electric field as a function of the aspect ratio. The beam is located 1 mm above the geometrical center. $E_v$ passes through zero at $a_0/b_0 = 1.7$. The blue curve aims to lead the eye.

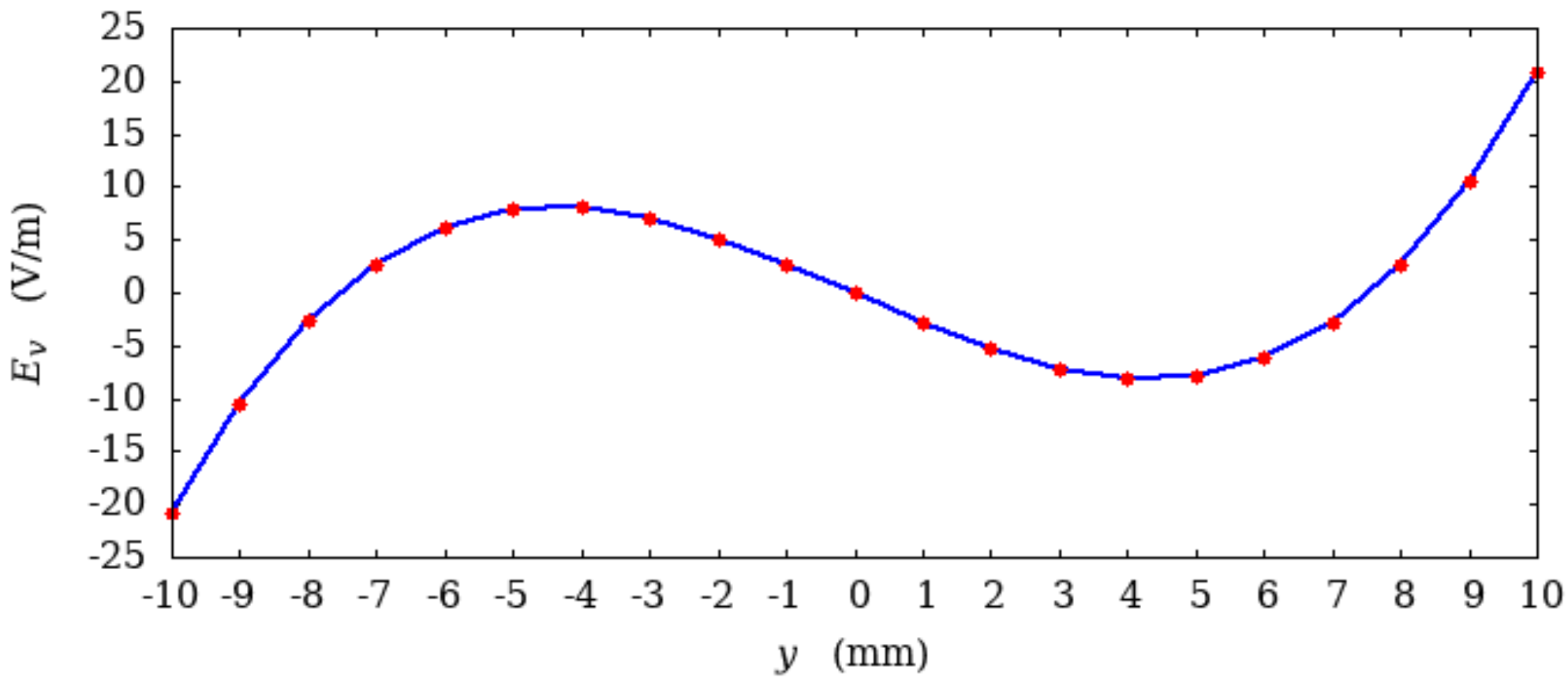

**Figure 5.** Net vertical electric field that is induced by the image beams, as a function of vertical beam offset $y$. The aspect ratio $a_0/b_0$ of the quadrupole aperture is 1.7, and the beam radius is 5 mm. The improvement with respect to the $a_0/b_0 = 1$ case is roughly two orders of magnitude.

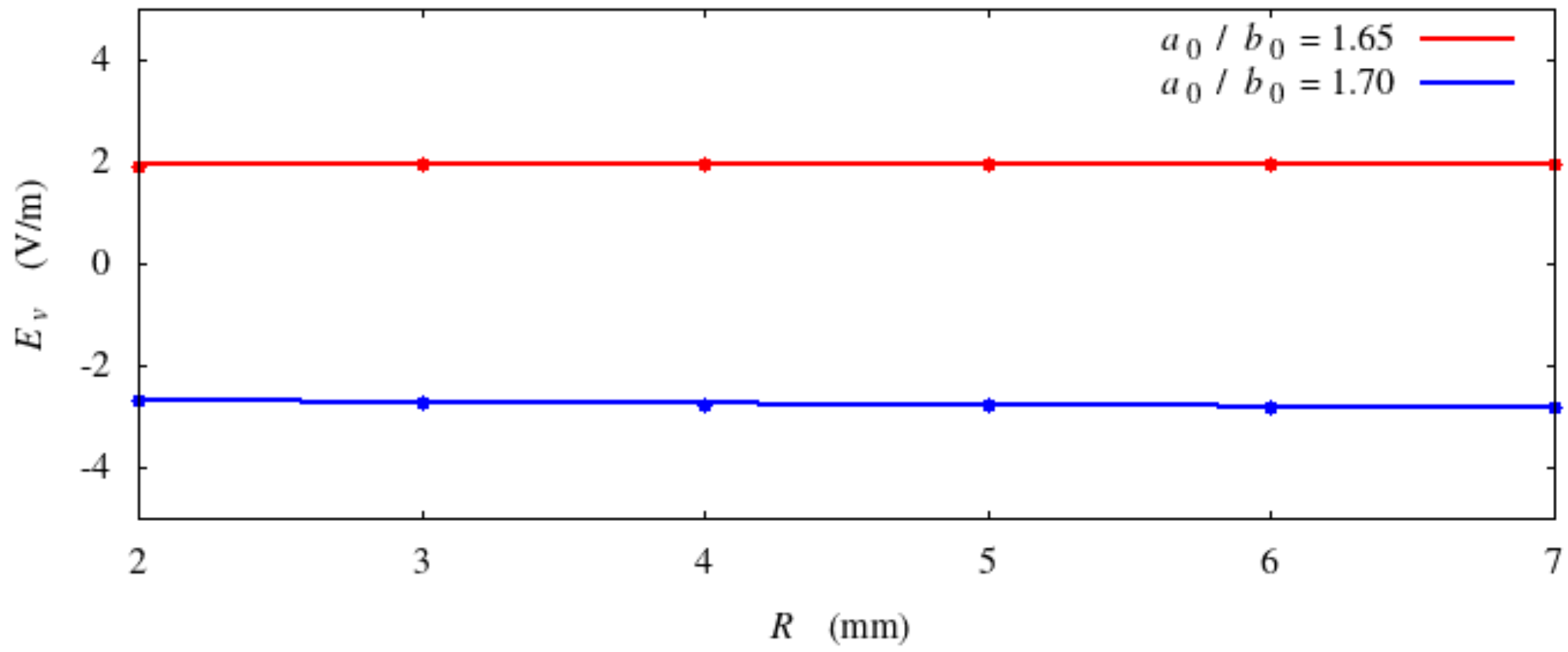

**Figure 6.** The vertical electric field on the beam is almost flat as a function of the beam radius. Comparison of the two plots implies that a better performance can be obtained with the aspect ratio of $a_0/b_0 = 1.67$.

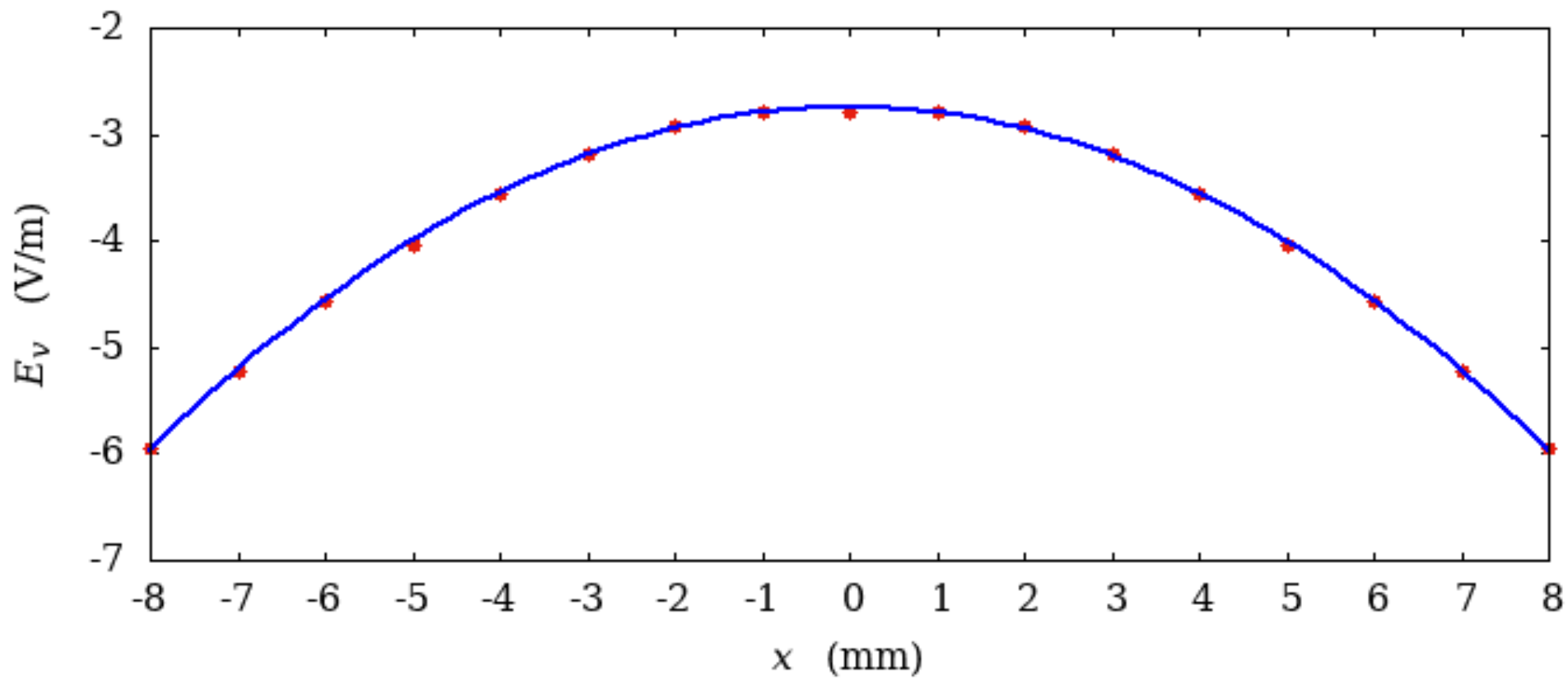

**Figure 7.** Vertical electric field as a function of horizontal offset. The simulations are conducted with $a_0/b_0 = 1.7$ and a vertical beam offset $y = 1$ mm. The increase in $E_v$ is insignificant, even though the effect does not cancel on opposite sides of $x = 0$.

Figure 6 shows $E_v$ as a function of radius for a beam located at $y = 1$ mm. The simulations are done with two different aspect ratios. In both cases, the vertical electric field is almost flat over the simulated beam radii. Comparison of the two plots implies that the method works better at $a_0/b_0 \approx 1.67$ if the vertical beam offset is around 1 mm.

Finally, horizontal beam offset is also investigated. These simulations are conducted with a beam of 5 mm radius and $y = 1$ mm vertical offset inside a quadrupole of $a_0/b_0 = 1.7$ aspect ratio. The $x$ range is within ±8 mm, which is comparable to the oscillation amplitude of the beam in the proton EDM experiment proposal. As shown in Figure 7, the vertical electric field for this range is growing as the beam moves away from the origin. Even though the left and right sides do not cancel each other, the horizontal oscillation can cause an increase in average $E_v$ only by a small fraction. Overall, the two orders of magnitude improvement is roughly conserved for all $x$ values.

## 4. Discussion and Conclusion

Even small electromagnetic fields can cause large systematic errors in spin physics experiments. In this work, it is shown that image charges at a normal electric quadrupole can result a non-negligible effect on the spin precession for the proton EDM experiment. One solution to this systematic error is optimizing the quadrupole profile. To our knowledge, this is the first study that provides numerical estimates of the effect, and how it changes with the aspect ratio of the quadrupole and other beam parameters.

According to the finite difference simulations, the vertical electric field on a beam inside a quadrupole can be reduced by approximately two orders of magnitude by rescaling quadrupole profile aspect ratio. Insignificant performance degradation with varying beam radii and horizontal offset is promising for usability in a variety of experimental cases. A 0.6 mm machining tolerance ensures a feasible manufacturing. In conclusion, this method becomes a useful tool for reducing image beam related vertical electric fields with no significant additional cost.

It is worth noting that application of the method is not limited to electric quadrupoles. A similar approach can be applied to electric sextupoles, as well. Moreover, in case the installation of vertical plates is not feasible, a grounded quadrupole profile can be installed inside other accelerator elements such as magnetic quadrupoles, magnetic sextupoles, magnetic deflectors, straight sections, and so on.

**Acknowledgment**

This work was supported by IBS-R017-D1 of the Republic of Korea. I would like to thank Yannis K. Semertzidis and William M. Morse for helpful discussions.